\documentclass[a4paper,11pt]{article}
\pdfoutput=1 

\usepackage{jinst} 
\usepackage{amsmath}
\usepackage{lineno}
\usepackage[symbol]{footmisc}

\usepackage{tabularx}
\title{Recent results from DEAP-3600}

\author[a,b,1]{M. Lai,\note{Corresponding author.}}

\affiliation[a]{Cagliari State University,\\Via Università 40, 09124 Cagliari, Italy}
\affiliation[b]{INFN Cagliari,\\S.P. Monserrato-Sestu Km 0.700, 09042 Monserrato, Italy}

\emailAdd{michela.lai@ca.infn.it}

\abstract{DEAP-3600 is the largest running dark matter detector filled with liquid argon, set at SNOLAB, in Sudbury, Canada, 2 km underground. The experiment holds the most stringent exclusion limit in argon for WIMPs above 20 GeV/c$^2$. In the most recent published analysis, the background events due to alpha-induced scintillation in the neck of the detector limited the sensitivity. The sensitivity of the detector in the next WIMP search will be improved thanks to the decrease in backgrounds achieved by hardware upgrades and applying multivariate analyses.
Moreover, the WIMP analysis has been revisited in terms of a non-relativistic effective field theory framework, and the impact of possible substructures in the galactic dark matter halo was explored. This analysis was motivated by the latest results from Gaia and the Sloan Sky Digital Survey. Here DEAP-3600 set the world's best exclusion limit for xenon-phobic dark matter scenarios. Finally, a custom-developed analysis has recently pointed out the extraordinary sensitivity to ultra-heavy, multi-scattering dark matter candidates, resulting in world-leading exclusion limits on two composite dark matter candidates up to Planck scale masses. These proceedings, after a quick overview of the dark matter detection in DEAP-3600, outline the detector upgrades and the dark matter search results from the collaboration of the last three years.}

\keywords{Noble liquid detectors, Dark Matter detectors, Scintillators, Cryogenic detectors}

\collaboration[c]{on behalf of DEAP-3600 collaboration}

\proceeding{LIght Detection In Noble Elements\\
  21-23 September 2022\\
 AstroCENT -Nicolaus Copernicus Astronomical Center}

\begin{document}

\maketitle

\flushbottom

\section{Dark matter detection in DEAP-3600}
\vspace{-2mm}
DEAP-3600 is a single-phase liquid argon detector designed to search for Weakly Interacting Massive Particles (WIMPs). The detector is located at SNOLAB, Sudbury, Ontario, 2 km underground (6 km water equivalent). During the second fill run, which started in November 2016 and ended in April 2020, an ultraviolet-absorbing acrylic vessel, consisting of a sphere with an 85 cm radius was filled up to 551 mm from the equator, resulting in a mass of 3279 $\pm$ 96 kg of atmospheric argon. The scintillation light from liquid argon is observed by 255 8" diameter Hamamatsu R5912 HQE photomultiplier tubes (PMTs) coupled to the vessel through 45 cm long acrylic light guides. The neck at the top of the vessel hosts the cooling coil and allows for operations into the vessel. The system is placed in a stainless steel shell, where the PMTs are set. High-density polyethylene and Styrofoam “filler blocks" are set between the acrylic light guides, allowing for the suppression of cosmogenic neutrons and for the PMTs to operate between 240 K and 290 K. The shell is submersed in a cylindric water tank filled with ultrapure water, which serves as muon veto together with the 48 outward-looking PMTs set on the external surface of the stainless-steel shell \cite{detector}. The light yield during this run has been essentially stable, with a variation as low as 0.3 PE/keV, as shown in Figure \ref{Figure: light yield}, with a slight decrease probably due to the instability of the water in the muon tank just after the refill. The next refill is scheduled for the end of 2023 after the completion of hardware upgrades.
\begin{figure}[htbp]
\centering 
\includegraphics[width=0.5\textwidth]{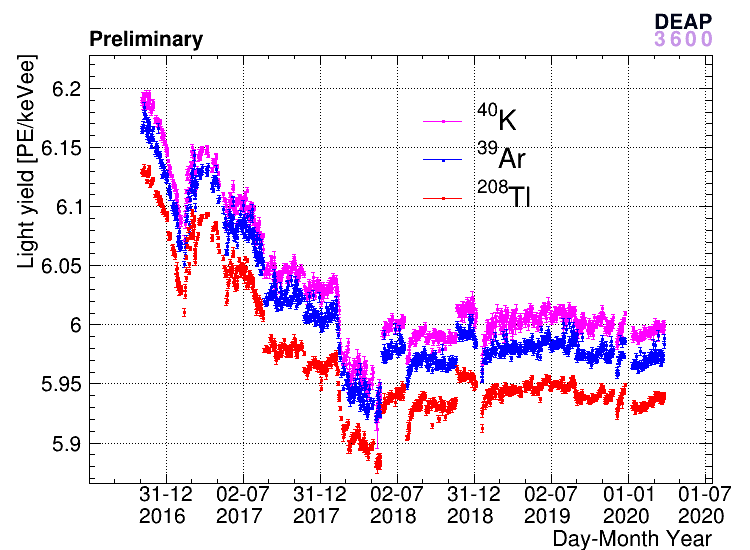}
\caption{\label{Figure: light yield} The stability of the light yield along the second fill was evaluated with three major components of the electron recoil background. It has been stable within 0.3 PE/keV, with a clear decrease in the first year, probably due to the warming up of the water tank just after the refill.}
\label{Figure: light yield}
\end{figure}   
\section{Background rejection}
\vspace{-2mm}
The main advantage of liquid argon as a target is the high discrimination level between nuclear and electron recoils, achievable with pulse shape discrimination (PSD). Specifically, the pulse shape in DEAP-3600 has been modeled as a singlet excited dimer with a time decay of 8 ns and a triplet state of 1.4 $\mu$s; in addition to these, a third term takes into account the time needed by an ionized electron to be captured back by an argon ion \cite{pulseshape}. Of the four PSD algorithms which were recently compared, the most efficient one was found to be F$_{prompt}$, defined as the fraction of the recorded scintillation light in the first 120 ns of the event, where the number of photoelectrons N$_{sc}$ is defined according to the Bayes' theorem, after removing the contribution from afterpulses. For events at 18 keV$_{ee}$, so about 110 N$_{sc}$, the leakage probability of electron recoils entering the nuclear recoil F$_{prompt}$ band is 10$^{-10}$ for a nuclear recoil acceptance of 50 \% \cite{psd}. 
The electron recoil background was fully modeled up to 8 MeV, including the $\beta$s and $\gamma$s released by radioactive traces in the inner detector, which mainly extends up to about 3 MeV \cite{em1}, and the neutron capture $\gamma$s (n, $\gamma$), which are the dominant background in the electron recoil band up to about 10 MeV \cite{em2}. \\
On the other hand, one of the most insidious backgrounds in the last dark matter search has been $\alpha$s from the neck of the detector. Once released by $^{210}$Po in the neck flowguides, these induce scintillation in the argon film condensed on the flowguides. Due to the geometry of the flowguides, the signal is partially shadowed and, if detected, can enter the WIMP Region of Interest (ROI) \cite{231d}. Neck $\alpha$s have been rejected by applying fiducial volume cuts based on the position reconstruction and the photon distribution along the PMTs and will be further mitigated with a recently developed multivariate analysis. 
Further suppression of neck $\alpha$s is expected after the ongoing hardware upgrades. New flowguides, coated with pyrene, a wavelength shifter, will be installed before the beginning of the third fill. Pyrene has two excitation states, with $\tau_m$ $\approx$ 250 ns and $\tau_e \approx$ 280 ns. Such a slow decay time allows for applying a powerful pulse shape discrimination of the neck $\alpha$s produced on the flowguides compared to the nuclear recoil events produced in the liquid argon itself. In addition, the internal cooling system will be replaced with an external one to prevent argon condensation on the flowguides \cite{pyrene}. 
An additional background that can be reconstructed in the WIMP ROI is the decay of $\alpha$s embedded in particulate dust in the liquid argon, which results in a degraded and isotropic scintillation signal. In the most recent simulations, this background has been modeled with the superposition of dust with different sizes, with a diameter from 1 $\mu$m to 50~$\mu$m. 
A removable pipe, going down to the bottom of the detector, has been designed to remove the dust from the vessel before the third fill. Once the detector is running, an external alternate cooling system will allow to periodically extract, filter, condense and refill the vessel through the neck to keep the dust amount at a negligible level.
\vspace{-2mm}
\section{Recent analysis}
\vspace{-2mm}
The last WIMP search for spin-independent interaction was published in 2019, where DEAP-3600 set the most stringent exclusion limit in argon for WIMP masses above 20 GeV/c$^2$, down to 3.9~$\times$~10$^{-45}$~cm$^2$  at 100 GeV/c$^2$ within 90 \% C.L. \cite{231d}. The same results have been interpreted within the context of the non-relativistic effective field theory (NREFT) framework \cite{halo}. 
The exclusion limits are shown in Figure \ref{Figure: halo2} for the operators giving a non-negligible interaction in argon. The limits are expressed in terms of the effective dark matter-proton cross-section $\sigma_p$ = $\frac{(c_i^p \mu_p)^2}{\pi}$, where $\mu_p$ is the dark matter-proton reduced mass. The dark matter-proton and neutron couplings are defined as 
\begin{equation}
    c_i^p = \frac{(c_i^0 + c_i^1)}{2}, ~~~~~~~~~~
    c_i^n = \frac{(c_i^0 - c_i^1)}{2};
\end{equation}
where c$_i^0$ and c$_i^1$ refer to the isoscalar and the isovector component of the coupling for the \textit{i}-th NREFT operator. The usual assumption in the WIMP searches corresponds to the dark matter-proton cross-section as evaluated for the O$_1$ operator and assuming c$_i^1 = 0$, which means  c$_i^n$ = c$_i^p$, so the isoscalar (IS) scenario. This assumption is the simplest one and allows for comparison among direct detection experiments based on different targets. Still, non-negligible differences arise in liquid argon when considering the operators above, not only assuming the isoscalar scenario but also the isovector one (IV, with c$_i^n$ = - c$_i^p$) and the xenon-phobic scenario (XP, for c$_i^n$ = - 0.7 c$_i^p$).
In addition, the exclusion limits from NREFT have been reported according to the halo's non-standard velocity distributions of the dark matter. Indeed, according to the second data release from the Gaia space mission \cite{gaia} and the data from the Sloan Digital Sky Survey (SDSS) \cite{ssds}, several kinematic substructures have been identified in the Milky Way halo.
Such observation suggests that a percentage of the dark matter $\eta_{sub}$ can also be in these substructures, in analogy with the baryonic matter, resulting in a deviation from the standard Maxwell-Boltzmann distribution truncated at the galactic escape velocity.
Here we focus on two examples: 
\begin{itemize}
    \item the Gaia Enceladus as modeled with a bimodal distribution comprising two Gaussian distributions. This substructure seems to have resulted from a merger event with a massive dwarf galaxy of about 5 $\times 10^{10}$ M$_{Sun}$, and shows a mean velocity lower than the average along the Milky Way. The dark matter fraction here is assumed to be in the range $\eta_{sub} = [0, 70]$ \%, to include all the suggested values up to today.
    \item the Group 1 (G1) of streams includes the Koppelman 1 stream and the retrograde infalling clumps (ICs) with a mean velocity of 400 km/s. Koppelman 1 is a stellar stream probably resulting from a recent accretion event in our galaxy. ICs are a model where dark matter accretes in the Milky Way beyond what is observed on stellar streams. These ICs can arise from past accretion events or intergalactic DM falling on our galaxy.
\end{itemize}
Figure \ref{Figure: halo2} shows the exclusion limits within the mentioned NREFT operators for a varying percentage $\eta_{sub}$ of dark matter in the G1 and the Gaia Enceladus kinematic substructures, respectively faster and slower than the rest of the Milky Way. In the bottom inset, the $\eta_{sub}$ is fixed to the value corresponding to the maximum of the exclusion plot for each operator and the given substructure to highlight the relative deviation from the standard halo model.
\begin{figure}[htbp]
\centering 
\includegraphics[width=0.9\textwidth]{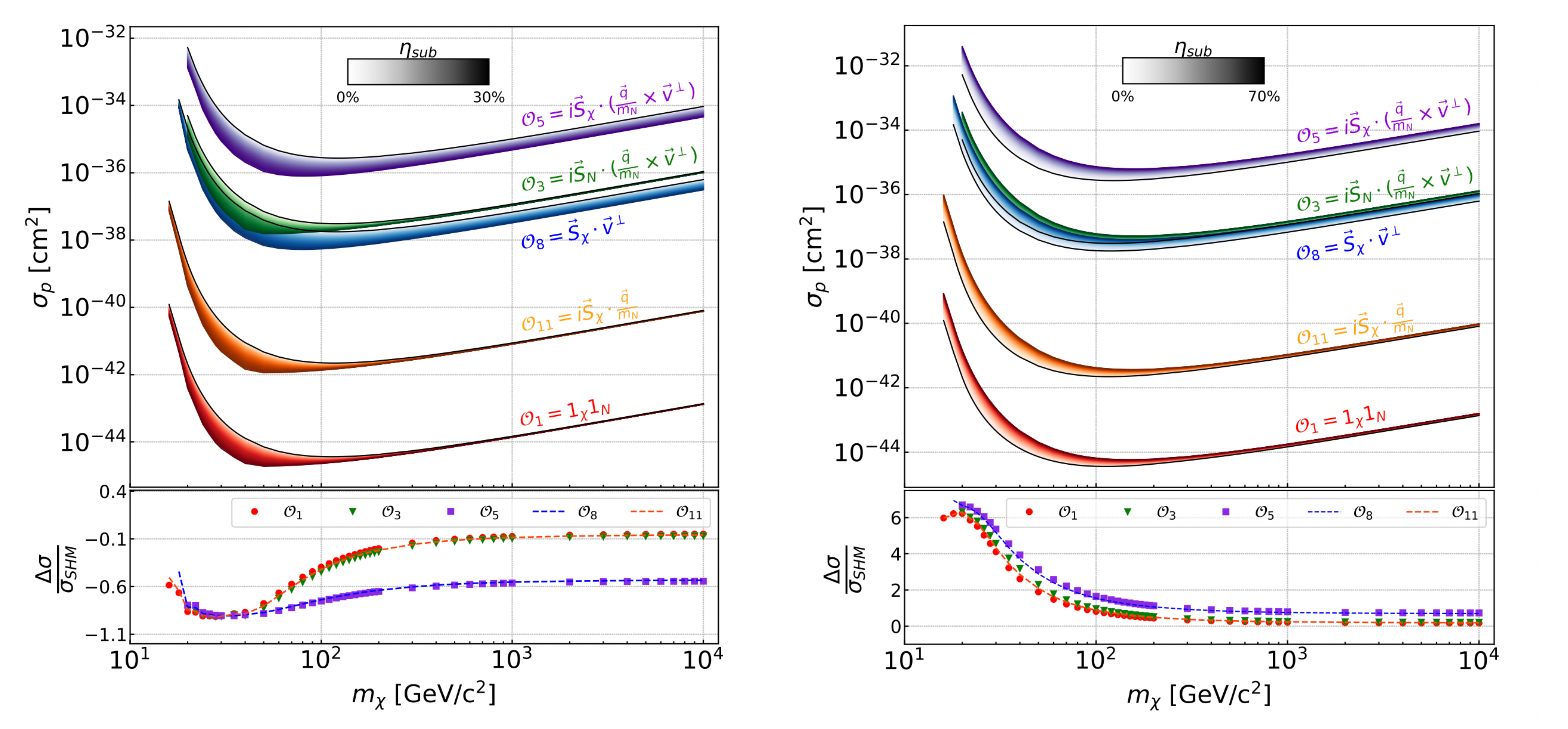}
\label{Figure: halo2}
\qquad
\caption{\label{Figure: halo2} Exclusion limits for all the NREFT operators, assuming isoscalar coupling, for the Group 1 class of substructures (Left) and the Gaia Enceladus (Right), for a varying percentage of dark matter in the substructures $\eta_{sub}$. The bottom inset shows the relative deviation of each operator for the given substructure at its maximum value compared to the standard halo model (SHM), with $\Delta \sigma$ = $\sigma_{sub} $- $\sigma_{SHM}$, assuming the higher value of the exclusion plot \cite{halo}. } 
\end{figure}
In the case of the G1, the dark matter candidate in the substructure would have higher kinetic energy and would be more likely to be detected. Hence the exclusion limits are strengthened as the $\eta_{sub}$ increases for G1 substructures, while the opposite behavior is observed for the Gaia Enceladus. The impact on the exclusion limits due to a comprehensive list of substructures was investigated in \cite{halo}.\\
Besides WIMPs, DEAP-3600 is also sensitive to ultra-heavy dark matter candidates at masses from 10$^7$ GeV to 10$^{19}$ GeV, which are expected in Grand Unification Theories and can be produced in out-of-equilibrium mechanisms in the early universe, such as decay products of the inflaton or as relics from a weakly-haired primordial black hole. The detector sensitivity is mainly limited by the low flux, which must be compensated by very high cross-sections in order to get a few events per year. Moreover, due to the high cross-section, the expected signal will be a track of nuclear recoils, essentially collinear due to the high ratio between the DM candidate mass and the argon nucleus one. An example of the photoelectron time distributions is reported in Figure \ref{Figure: mimp} for a simulated candidate at mass $m_\chi = 10^{18}$ GeV/c$^2$ and DM-argon nucleus cross-section of $\sigma_{T-\chi}=$2.0~$\times$ 10$^{-21}$~cm$^2$.
N$_{peaks}$ reports the number of outstanding peaks along the photoelectron distribution according to a derivative-based algorithm. Both F$_{prompt}$ and N$_{peaks}$ decrease as the cross-section increases. Indeed, the increasing number of scatters determines a higher amount of pulses after the prompt time window of 120 ns and a more likely merging of the pulses, which start to be unlikely to be dominant with respect to the previous ones. The dominant background up to 10 MeV are pile-up events due to two or more electron recoils happening in the same acquisition window, which would give an F$_{prompt}$ around 0.3 or less and N$_{peaks} > $ 1, eventually overlapping with the signal. 
Given the knowledge of the single-scatter electromagnetic background, \cite{em1}, a specific threshold on N$_{peaks}$ was set according to the specific energy range to reject pile-up events. The selection cut on N$_{peaks}$ was applied up to 10 MeV of deposited energy, where the simulations could be validated, determining three Regions of Interest (ROIs), as shown in Table \ref{tab:roibkgds}. ROI4 extends above 10 MeV up to the ceiling of the analysis, set at about 60 GeV; here, no N$_{peaks}$ cut is applied. A muon veto cut is extended along all four ROIs, rejecting any event happening within [-10, 90] $\mu$s from the muon veto trigger. After the unblinding of three years of data, with a resulting lifetime of (813 $\pm$ 8) live-days and a total background level of (0.05 $\pm$ 0.03) events,  no event was found in any of the four ROIs. 
\begin{table*}[htb]
\begin{center}
\begin{small}
 \begin{tabular}{cllcccc} 
 \hline\hline 
  ROI & PE range &  Energy [MeV]  & $\text{N}_\text{peaks}^\text{min}$ & $\text{F}_\text{prompt}^\text{max}$ &  $\mu_b$ & $\text{N}_\text{obs.}$ \\ \hline
 1 & 4000-20000  & 0.5-2.9  & 7  & 0.10  & (4 $\pm$ 3) $\times$ 10$^{-2}$  & 0 \\
 2 & 20000-30000  & 2.9-4.4  & 5  & 0.10  & (6 $\pm$ 1) $\times$ 10$^{-4}$ & 0 \\  
 3 & 30000-70000& 4.4-10.4 & 4 & 0.10 & (6 $\pm$ 2) $\times$ 10$^{-4}$ & 0 \\
 4 & 70000-4 $\times$ 10$^8$ & 10.4-60000 & 0 & 0.05 & (10 $\pm$ 3) $\times$ 10$^{-3}$ & 0 \\ 
 \hline\hline
\end{tabular}
\caption{ROI definitions according to the PE and energy range, the threshold on the number of dominant peaks N$_{peaks}$ and the upper selection cut on F$_{prompt}$, as well as the background expectations $\mu_b$ and observed event counts $\text{N}_\text{obs.}$ after the unblinding with 813 days of exposure.}
 \label{tab:roibkgds}
 \end{small}
\end{center}
\end{table*}
\vspace{-2mm}
The exclusion limits are set for two composite dark matter models, of which only Model 2, referring to dark matter "nuggets," is reported in Figure~\ref{Figure: mimp}.
As validated simulations were available for candidates with cross-sections giving a signal extending up to about 10 MeV of deposited energy, above this point, the acceptance was conservatively set at 35 \%, giving the extrapolated exclusion limits in the grey area in Figure \ref{Figure: mimp}. 
\begin{figure}[htbp]
\centering 
\includegraphics[width=0.9\textwidth]{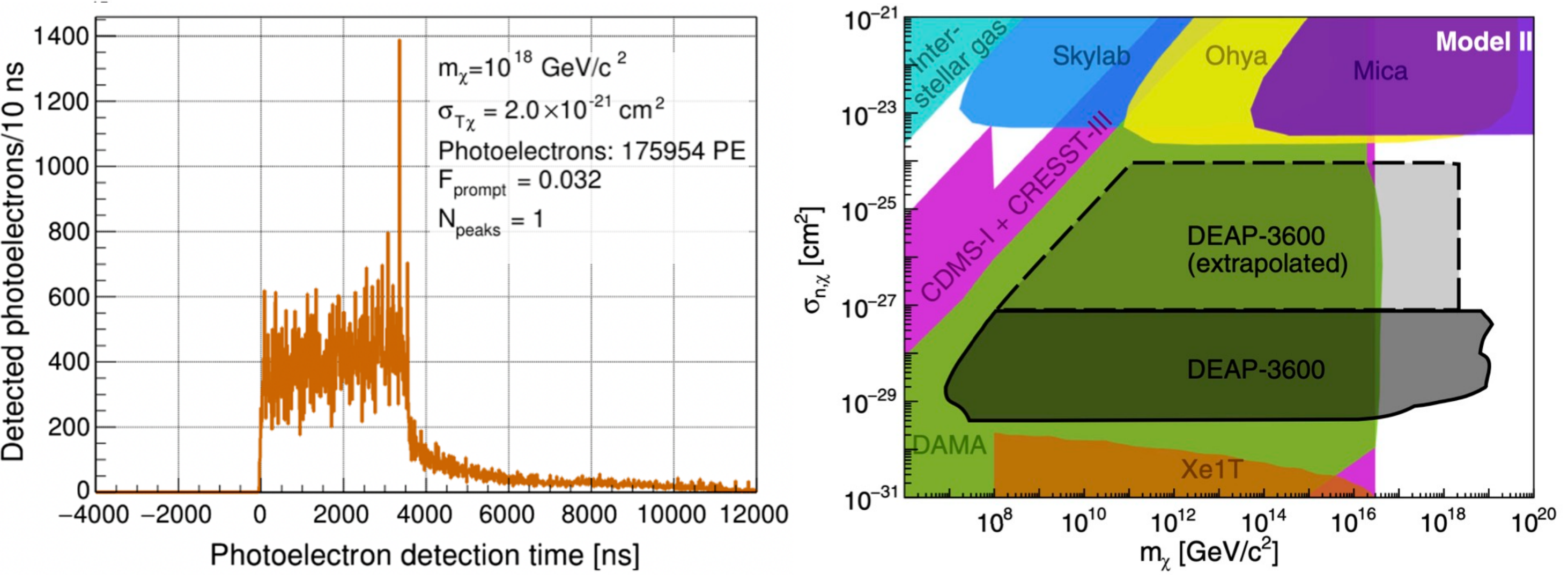}
\label{Figure: mimp}
\qquad
\caption{\label{Figure: mimp} Left: Photoelectron time distribution for a heavy, multi-scatter dark matter candidate. Right: Exclusion limits set for dark nugget candidates, including the ones extrapolated in ROI4, by assuming a conservative acceptance of 35 \% \cite{mimp}.}
\end{figure}
Thanks to the high exposure and the custom-developed analysis, DEAP-3600 is the first running experiment that could exclude dark matter up to Planck scale masses below 10$^{-24}$ cm$^2$.
\vspace{-2mm}
\section{Conclusion}
\vspace{-2mm}
DEAP-3600 has shown an extraordinary sensitivity to dark matter candidates, thanks to its exposure and the low background level achieved with its liquid argon target. While hardware and analysis upgrades will guarantee an even lower background level in the next detector refill, scheduled for the end of 2023, the experiment still holds the most stringent exclusion limit in liquid argon for WIMPs above 20 GeV/c$^2$, as well as the world-leading exclusion limit for xenon-phobic couplings above 100 GeV/c$^2$. Moreover, its exposure and the custom analysis for multi-scattering dark matter brought the first exclusion limits for composite dark matter candidates up to Planck scale masses. 
\vspace{-2mm}
\section*{Acknowledgments}
\vspace{-2mm}
We thank the Natural Sciences and Engineering Research Council of Canada,
the Canadian Foundation for Innovation (CFI),
the Ontario Ministry of Research and Innovation (MRI), 
and Alberta Advanced Education and Technology (ASRIP),
Queen's University,
the University of Alberta,
Carleton University,
the Canada First Research Excellence Fund,
the Arthur B.~McDonald Canadian Astroparticle Research Institute,
DGAPA-UNAM (PAPIIT No.~IN108020) and Consejo Nacional de Ciencia y Tecnolog\'ia (CONACyT, Mexico, Grant A1-S-8960),
the European Research Council Project (ERC StG 279980),
the UK Science and Technology Facilities Council (STFC) (ST/K002570/1 and ST/R002908/1),
the Leverhulme Trust (ECF-20130496),
the Russian Science Foundation (Grant No. 21-72-10065),
the Spanish Ministry of Science and Innovation (PID2019-109374GB-I00), 
and the International Research Agenda Programme AstroCeNT (MAB/2018/7)
funded by the Foundation for Polish Science (FNP) from the European Regional Development Fund.
Studentship support from
the Rutherford Appleton Laboratory Particle Physics Division,
STFC and SEPNet Ph.D. is acknowledged.
We thank SNOLAB and its staff for support through underground space, logistical, and technical services.
SNOLAB operations are supported by the CFI
and Province of Ontario MRI,
with underground access provided by Vale at the Creighton mine site.
We thank Vale for their continuing support, including the work of shipping the acrylic vessel underground.
We gratefully acknowledge the support of Compute Canada,
Calcul Qu\'ebec,
the Centre for Advanced Computing at Queen's University,
and the Computation Centre for Particle and Astrophysics (C2PAP) at the Leibniz Supercomputer Centre (LRZ)
for providing the computing resources required to undertake this work.

\end{document}